# A Novel Algebraic System in Quantum Field Theory

A. D. Alhaidari[a] and A. Laradji [b]

[a] *Saudi Center for Theoretical Physics, P.O. Box 32741, Jeddah 21438, Saudi Arabia*

[b] *Mathematics Department, King Fahd University of Petroleum & Minerals, Dhahran 31261, Saudi Arabia*

**Abstract**: An algebraic system is introduced, which is very useful for doing scattering calculations in quantum field theory. It is the set of all real numbers greater than or equal to $-m^2$ with parity designation and a special rule for addition and subtraction, where $m$ is the rest mass of the scattered particle.

## 1. Introduction and formulation

In physics, quantum field theory (QFT) was developed to describe structureless elementary particles (e.g., electrons, quarks, photons, etc.), their interaction with each other and with their environment [1-3]. A graphical technique to account for physical processes in QFT (e.g., scattering) is by using diagrams known as the Feynman diagrams [4]. These consist of points (vertices) connected by lines (propagators) [5,6]. The lines represent free propagation of elementary particles and the points represent the interaction among particles meeting at those points. On the other hand, a QFT for particles with structure has recently been proposed by the first author [7]. In addition to the usual formulation of conventional QFT, the new theory relies predominantly on the properties of orthogonal polynomials. In the remainder of this introductory section, we summarize the formulation of this theory and in Section 2 we address scattering in the theory and show how a novel algebraic system appears naturally. In Section 3, we give a rigorous mathematical definition of this algebraic system and in Section 4 we demonstrate how scattering calculation is carried out in the new theory using Feynman diagrams and utilizing the novel algebraic system. Finally, we end with a brief summary and conclusion in Section 5.

One way to represent free scalar particles in QFT is to utilize solutions of the Klein-Gordon wave equation in (3+1)-dimensional Minkowski space-time that reads

$$(\partial_t^2 - \vec{\nabla}^2 + m^2)\Psi(t,\vec{r}) = 0, \quad (1)$$

where $m$ is the rest mass of the scalar particle and we have adopted the relativistic units $\hbar = c = 1$. Recently, a formulation of QFT for elementary particles that have internal structure was developed in which the quantum field operator $\Psi(t,\vec{r})$ is written as a Fourier expansion over the energy domain consisting of continuous and discrete components:

$$\Psi(t,\vec{r}) = \int_\Omega e^{-iEt}\psi(E,\vec{r})a(E)dE + \sum_{j=0}^{N} e^{-iE_j t}\psi_j(\vec{r})a_j. \quad (2)$$

The integral over $\Omega$ represents the continuous energy spectrum of the particle whereas the sum represents its structure, which is resolved in the energy and of size $N+1$. This formulation of QFT is referred to by the acronym SAQFT that stands for "Structural Algebraic QFT" and could be useful in treating elementary particles that are thought to be structureless at low energy scale [7]. We take $\Omega$ to stand for the single energy interval $E^2 \geq m^2$ and take $0 \leq E_j^2 < m^2$. The objects $a(E)$ and $a_j$ are field operators (the vacuum annihilation operators) that satisfy the conventional commutation relations: $[a(E), a^\dagger(E')] = \delta(E - E')$ and $[a_i, a_j^\dagger] = \delta_{i,j}$.

Now, the continuous and discrete energy kernels in the Fourier expansion (2) have distinct properties but both are pointwise finite in space. The discrete kernel $\psi_j(\vec{r})$ is confined and vanishes asymptotically whereas the continuous kernel $\psi(E, \vec{r})$ remains asymmetrically finite. Nonetheless, we can find a complete set of basis functions in space where we can expand both kernels using elements of the same basis but with different types of expansion coefficients. If such a basis set is designated as $\{\phi_n(\vec{r})\}$, then we can write these Fourier energy components as the following pointwise convergent series

$$\psi(E, \vec{r}) = \sum_{n=0}^{\infty} A_n(E)\phi_n(\vec{r}) := A_0(E) \sum_{n=0}^{\infty} p_n(z)\phi_n(\vec{r}), \tag{3a}$$

$$\psi_j(\vec{r}) = \sum_{n=0}^{\infty} B_n(E_j)\phi_n(\vec{r}) := B_0(E_j) \sum_{n=0}^{\infty} p_n(z_j)\phi_n(\vec{r}), \tag{3b}$$

where $z$ is some proper function of the energy called the spectral parameter, which is to be determined, and $p_0(z) = 1$. For scalar particles, we require that $\{\phi_n(\vec{r})\}$ satisfy the following differential relation

$$-\vec{\nabla}^2 \phi_n(\vec{r}) = \alpha_n \phi_n(\vec{r}) + \beta_{n-1} \phi_{n-1}(\vec{r}) + \beta_n \phi_{n+1}(\vec{r}), \tag{4}$$

where $\{\alpha_n, \beta_n\}$ are real constants such that $\beta_n \neq 0$ for all $n$. Using (4) in the free Klein-Gordon wave equation (1) gives the following algebraic relation

$$zp_n(z) = \alpha_n p_n(z) + \beta_{n-1} p_{n-1}(z) + \beta_n p_{n+1}(z), \tag{5}$$

for $n = 1,2,3,\ldots$ and giving $z = E^2 - m^2$, $z_j = E_j^2 - m^2$. This is a symmetric three-term recursion relation that makes $\{p_n(z)\}$ a sequence of polynomials in $z$ with the two initial values $p_0(z) = 1$ and $p_1(z) = \frac{z-\alpha_0}{\beta_0}$. On the other hand, for spinor particles and due to the multiplicity of the quantum field components, the basis consists of a number of sets equals to the number of field components. For example, in 3+1 space-time, the spinor quantum field $\Psi^{\uparrow\downarrow}(t, \vec{r})$ is a 4-component field and the 2-component basis sets $\{\phi_n^\pm(\vec{r})\}$ are required to satisfy

$$-i\vec{\sigma} \cdot \vec{\nabla} \phi_n^-(\vec{r}) = c_n \phi_n^+(\vec{r}) + d_n \phi_{n-1}^+(\vec{r}), \tag{6a}$$

$$-i\vec{\sigma} \cdot \vec{\nabla} \phi_n^+(\vec{r}) = c_n \phi_n^-(\vec{r}) + d_{n+1} \phi_{n+1}^-(\vec{r}), \tag{6b}$$

where $\{\vec{\sigma}\}$ are the three $2 \times 2$ Pauli spin matrices and $\{c_n, d_n\}$ are real constant parameters. Using (6) in the coupled 4-component Dirac equation, we also obtain the three-term recursion relation (5) but for $\{p_n^{\uparrow\downarrow}(z)\}$ corresponding to two sets of recursion coefficients $\{\alpha_n^{\uparrow\downarrow}, \beta_n^{\uparrow\downarrow}\}$ that depend differently on the constants $\{c_n, d_n\}$. For more details on the spinor formulation in SAQFT, interested readers are referred to Appendix B in Ref. [7].

Favard theorem [8] (a.k.a. Shohat-Favard theorem [9] or the spectral theorem [10]) states that a sequence of polynomials satisfying the three-term recursion relation (5) with $\beta_n \neq 0$ for all $n$ is an orthogonal and a complete set. In general, the polynomial solution of Eq. (5) satisfies the following orthogonality relation [10-12]

$$\int_\Omega \rho(z) p_n(z) p_m(z) dz + \sum_{j=0}^{N} \xi(z_j) p_n(z_j) p_m(z_j) = \delta_{n,m}, \tag{7}$$

where $\rho(z)$ is the continuous component of the weight function and $\xi(z_j)$ is the discrete component. These weight functions are positive definite and related to the energy functions $A_0(E)$ and $B_0(E)$ as $A_0^2(E)dE = \rho(z)dz$ (with $\frac{dz}{dE} > 0$ for $E \in \Omega$) and $B_0^2(E_j) = \xi(z_j)$.

It should be clear from the above analysis that the wave equation (1), which is the Klein-Gordon equation, is equivalent to the three-term recursion relation (5). Now, since in conventional QFT, the solution of Eq. (1) is used in defining scalar particles, so too is the polynomials solution of the recursion (5) in SAQFT. Moreover, as shown briefly above and detailed in Ref. [7], the Dirac equation is equivalent to another three-term recursion relation whose solution is the alternative set of orthogonal polynomials $\{p_n^{\uparrow\downarrow}(z)\}$. Therefore, these alternative polynomials are associated with spinor particles. In-as-much as conventional QFT distinguishes particles from one another by assigning different properties to the solutions of the wave equation (Klein-Gordon or Dirac), so too does SAQFT by assigning different properties to the associated orthogonal polynomials. For example, one can give different values to the physical parameters that appear in the recursion coefficients $\{\alpha_n, \beta_n\}$ or in the spectral parameter $z$. In Ref. [7], we propose that the Wilson polynomial $W^\mu(z; a, b, c)$ with $-3 < \mu < -2$ can be used to represent baryons whereas the continuous dual Hahn polynomials $S^\mu(z; a, b)$ with $-2 < \mu < -1$ can be used for mesons. The parameters $\{a, b, c\}$ could assume one of six values corresponding to one of the six flavors of the constituent quarks whereas their conjugates represent the anti-quarks.

## 2. Scattering in SAQFT

In conventional QFT, the propagators in the Feynman diagrams are tagged with the energy-momentum four-vector $(E, \vec{k})$. However, in SAQFT these propagators are tagged with the spectral parameter $z$. For free scalar particles, $E^2 > m^2$ making $z$ positive. However, doing scattering calculation with the Feynman diagrams in closed loops, one should integrate and sum over all possible values of the real energy in $\Omega \cup \{E_j\}_{j=0}^N$ (i.e., $E^2 \geq 0$) making the values of these spectral parameters $z$ greater than or equal to $-m^2$. At each vertex in the Feynman diagrams, the energy-momentum 4-vector is conserved. For example, when calculating the first order correction to the three-particle interaction vertex, we encounter loop diagrams similar to that shown in Figure 1 where the spectral parameters $\{x, x', y, y', z, u\}$ are indicated on their respective propagators. Choosing a counterclockwise loop, the three energy conservation equations are: $E(x) = E(u) - E(x')$, $E(y) = E(y') - E(u)$, and $E(z) = E(y') - E(x') = E(x) + E(y)$, where $E(a) = \pm\sqrt{a + m^2}$. This leads to a special rule for adding and subtracting spectral parameters. For example, $E(z) = E(x) + E(y)$ gives

$$z = x + y + m^2 + 2\,sgn\,\sqrt{(x + m^2)(y + m^2)}, \tag{8}$$

where "sgn" is $\pm$, which is the product of the signs of the two energies $E(x)$ and $E(y)$. Moreover, the sign of the energy $E(z)$ is the sign of $E(x) + E(y)$. Therefore, for the spectral parameters to contain full physical information, they must carry the sign of their corresponding energies. Hence, we associate with each spectral parameter a $\pm$ parity, which will be indicated as superscript on the parameter. For example, the parity of $z$ in (8) is the sign of $\sigma\sqrt{x + m^2} + \tau\sqrt{y + m^2}$ where $\sigma$ is the sign of the energy $E(x)$ (i.e., parity of $x$) and $\tau$ is the sign of the energy $E(y)$ (i.e., parity of $y$). Thus, we rewrite (8) properly as follows

$$z^\rho = \left(x + y + m^2 + 2\sigma\tau\sqrt{(x + m^2)(y + m^2)}\right)^\rho := x^\sigma \oplus y^\tau, \tag{9}$$

where $\rho$ is the (nonzero) sign of the energy $E(z)$ (i.e., parity of $z$). This equation defines the operation of addition of the spectral parameters. Repeating the same for the energy conservation equation $E(z) = E(y') - E(x')$, we obtain the following rule for the subtraction of spectral parameters

$$z^\rho = \left(y' + x' + m^2 - 2\tau\sigma\sqrt{(y' + m^2)(x' + m^2)}\right)^\rho := y'^\tau \ominus x'^\sigma. \tag{10}$$

where $\rho$ is the (nonzero) sign of $\tau\sqrt{y' + m^2} - \sigma\sqrt{x' + m^2}$. The parity of each spectral parameter $\{x, x', y, y', z, u\}$ in the figure (not shown) is the sign of the corresponding energy.

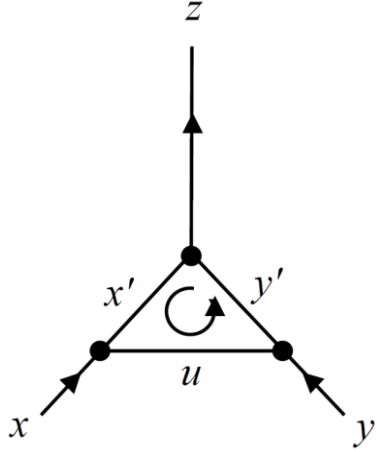

**Fig.1**: One of six loops in the Feynman diagrams used for calculating the first order correction to the interaction vertex.

Interested readers are referred to Ref. [7] for further details on the scattering calculation in SAQFT that utilizes this algebraic system. In the following section, we make a proper and rigorous mathematical definition of the underlying algebraic structure emerging from the physics of scattering with scalar particles in SAQFT as outlined above and detailed in Ref. [7].

## 3. The algebraic system

The novel algebraic system emerging from the physical application in QFT presented in the previous section and exhibited by the addition and subtraction rules of the spectral parameters as shown by Eq. (9) and (10) will now be given a proper mathematical definition.

**Notation**. Fix $r$ in $\mathbb{R}$ and let $\mathbb{R}^r = ([r, \infty) \times \{-1, 1\}) \setminus (r, -1)$, where the pair $(x, \sigma)$ is denoted by $x^\sigma$. For each $z \in \mathbb{R}$, let $s(z)$ be the (nonzero) sign of $z$: $s(z) = 1$ if $z \geq 0$, $s(z) = -1$ if $z < 0$.

Define operations $\oplus$ and $\otimes$ on $\mathbb{R}^r$ by

$$x^\sigma \oplus y^\tau = \left(x + y - r + 2\sigma\tau\sqrt{(x-r)(y-r)}\right)^{s(z)}, \tag{11}$$

$$x^\sigma \otimes y^\tau = ((x-r)(y-r) + r)^{s(w)}, \tag{12}$$

where $z = \sigma\sqrt{x - r} + \tau\sqrt{y - r}$ and $w = \sigma\tau\sqrt{(x - r)(y - r)}$.

–4–

**Proposition.** With the above notation, $(\mathbb{R}^r,\oplus,\otimes)$ is a field isomorphic to the usual field of real numbers. In particular, $(\mathbb{R}^r,\oplus)$ and $(\mathbb{R}^r\setminus\{(r,1)\},\otimes)$ are Abelian groups.

**Proof.** Let $f\colon\mathbb{R}^r\to\mathbb{R}$ and $g\colon\mathbb{R}\to\mathbb{R}^r$ be the functions defined respectively by

$$f(x^\sigma) = \sigma\sqrt{x-r}, \tag{13}$$
$$g(x) = (x^2+r)^{s(x)}. \tag{14}$$

A routine verification shows that $f$ and $g$ are inverses of each other and that

$$x^\sigma \oplus y^\tau = g(f(x^\sigma)+f(y^\tau)). \tag{15}$$
$$x^\sigma \otimes y^\tau = g(f(x^\sigma)f(y^\tau)). \tag{16}$$

Hence $f$ is a bijection that preserves the operations, and therefore $(\mathbb{R}^r,\oplus,\otimes)$ is a field isomorphic to the field $(\mathbb{R},+,\times)$ under the function $f$. The last statement follows from the fact that the additive identity of $\mathbb{R}^r$ is $g(0)$, i.e. $(r,1)$ (or, in our notation, $r^1$). ∎

Note that the multiplicative identity of $\mathbb{R}^r$ is $g(1)$, i.e. $(1+r,1)$, and that the additive and multiplicative inverses of $x^\sigma$ are $g(-f(x^\sigma))$ and $g\left(\frac{1}{f(x^\sigma)}\right)$ (for $x\neq r$), respectively, i.e. $x^{s(-\sigma\sqrt{x-r})}$ and $\left(\frac{1}{x-r}+r\right)^\sigma$.

If we denote subtraction on $\mathbb{R}^r$ by $\ominus$, i.e. $x^\sigma \ominus y^\tau = g(f(x^\sigma)-f(y^\tau))$, then

$$x^\sigma \ominus y^\tau = \left(x+y-r-2\sigma\tau\sqrt{(x-r)(y-r)}\right)^{s(z)}, \tag{17a}$$

where $z = \sigma\sqrt{x-r}-\tau\sqrt{y-r}$, so that

$$y^\tau \ominus x^\sigma = \left(x+y-r-2\sigma\tau\sqrt{(x-r)(y-r)}\right)^{s(-z)}. \tag{17b}$$

(note that $\ominus$ is not commutative since $x^\sigma \ominus y^\tau$ and $y^\tau \ominus x^\sigma$ may have different parities).

Other properties of $\mathbb{R}$ are also inherited by $\mathbb{R}^r$ via the bijection $f$ above. For example, $(\mathbb{R}^r, d)$ where $d(x^\sigma,y^\tau) = |f(x^\sigma)-f(y^\tau)|$ is a metric space.

As shown in Section 1 and detailed in Ref. [7], this algebraic system is very useful in relativistic scattering calculations using Feynman diagrams in SAQFT if we take the real constant $r = -m^2$. In that case, $f(x^\sigma) = \sigma\sqrt{x+m^2}$ becomes the relativistic energy of the particle associated with the spectral parameter $x^\sigma$. Finally, it is worth noting that $\mathbb{R}^r$ together with addition $\oplus$ and a scalar multiplication $*$ defined by $a*x^\sigma = (a^2(x-r)+r)^{s(a\sigma\sqrt{x-r})}$, for each real number $a$, will turn $\mathbb{R}^r$ into a real vector space. This extra structure could have a physical interpretation and might be useful for certain applications in QFT.

## 4. Scattering example

As an illustration, we give an example where we show how to evaluate the Feynman diagrams occurring in scalar SAQFT by utilizing the algebraic structure introduced here. Let us consider a physical system with the nonlinear self-interaction term $g|\Psi(t,\vec{r})|^3$ where $g$ is a coupling parameter. Therefore, Fig. 1 is one such diagram, which is used in the calculation that contributes to the 3rd order correction (i.e., up to $g^3$) of the interaction vertex. Since scalar particles in SAQFT are fully described by the properties of the orthogonal polynomials $p_n(z)$ defined by Eq. (5) and Eq. (7) along with their initial values $p_0(z)$ and $p_1(z)$, the corresponding propagators in the Feynman diagrams will be labeled by the polynomial index and its argument (the spectral parameter). Therefore, for calculation purposes we should include in Fig. 1 the missing polynomial index on each propagator along with the spectral parameter. Let us choose the indices $\{i,j,k,l,m,n\}$ for the propagators that correspond to the spectral parameters $\{x',y',u,x,y,z\}$, respectively. Since in scattering experiments, the input/output channels are selected, the two sets $\{l,m,n\}$ and $\{x,y,z\}$ are fixed. However, each element in the set $\{i,j,k\}$ runs from 0 to $\infty$ whereas each element in the set $\{x',y',u\}$ goes over the entire range of the continuous and discrete spectra from $-m^2$ to $\infty$. That is, they span $\Omega$ and run over $\{E_j\}$ from $j=0$ to $j=N$. However, energy conservation dictates that $z^\rho = x^\sigma \oplus y^\tau$, $x'^{\sigma'} = u^\zeta \ominus x^\sigma$ and $y'^{\tau'} = u^\zeta \oplus y^\tau$ leaving $u^\zeta$ as the only arbitrary spectral parameter. Consequently, if the bare interaction vertex is $g\{\eta_n^{m,k}\}$, then the Feynman diagram of Fig. 1 results in the following infinite sum (see Section 4 of Ref. [7])

$$\frac{g^3}{3!}\rho(x')\rho(y')\rho(u)\sum_{i,j,k=0}^{\infty} \eta_n^{i,j}\eta_l^{i,k}\eta_m^{j,k} p_i^2(x')p_j^2(y')p_k^2(u)$$

$$= \frac{g^3}{3!}\rho(u \ominus x)\rho(u \oplus y)\rho(u)\sum_{i,j,k=0}^{\infty} \eta_n^{i,j}\eta_l^{i,k}\eta_m^{j,k} p_i^2(u \ominus x)p_j^2(u \oplus y)p_k^2(u) \quad (18)$$

where each term is to be integrated and summed over the continuous and discrete energy spectra associated with the spectral parameter $u$ as follows

$$\int \rho(u \ominus x)\rho(u \oplus y)\rho(u) p_i^2(u \ominus x)p_j^2(u \oplus y)p_k^2(u) du$$
$$+ \sum_{t=0}^{N} \xi(u_t \ominus x)\xi(u_t \oplus y)\xi(u_t) p_i^2(u_t \ominus x)p_j^2(u_t \oplus y)p_k^2(u_t) \quad (19)$$

We have removed the parity designation from the spectral parameters in (18) and (19) for simplicity of the notation. It was shown in Ref. [7] that the integral (19) is one of the fundamental SAQFT integrals, which is finite and goes to zero fast enough as the indices $\{i,j,k\}$ go to infinity. To account for the full 3rd order correction to the interaction vertex, one should also add five more diagrams in addition to that shown in Fig. 1; each with a single loop. The details of this calculation are given in Section 4 of Ref. [7].

## 5. Conclusion

A formulation of QFT for elementary particles with structure has recently been proposed and referred to by the acronym SAQFT. Doing scattering calculation in the theory using Feynman diagrams along with the energy-momentum conservation forced us to employ a novel algebraic system in the spectral parameter space. In this short article, we gave a rigorous mathematical definition of this algebraic system and showed how it comes about by giving a simple scattering example. We focused our development on the scalar formulation of SAQFT, however, the spinor and vector formulation follow a similar framework but with multiple components.